# Crystal Growth of a New 8H Perovskite $Sr_8Os_{6.3}O_{24}$ Exhibiting High $T_C$ Ferromagnetism

*Gohil S. Thakur [1,2], Thomas Doert [2], Shrikant Mohitkar [1], Walter Schnelle[1], Claudia Felser [1], and Martin Jansen [1,3]\**

[1] Max Planck Institute for Chemical Physics of Solids, Nöthnitzer Str. 40, 01187 Dresden, Germany

[2] Faculty of Chemistry and Food Chemistry, Technical University, 01069 Dresden, Germany

[3] Max Planck Institute for Solids State Research, Heisenberg Str. 1, 70569 Stuttgart, Germany

**ABSTRACT**: Single crystals of a new twinned hexagonal perovskite compound $Sr_8Os_{6.3}O_{24}$ have been synthesized, and structural and magnetic properties have been determined. The compound crystallizes in a hexagonal cell with lattice parameters $a$ = 9.6988(3) Å and $c$ = 18.1657(5) Å. The structure is an eight-layered hexagonal B-site deficient perovskite with the layer sequence $(ccch)_2$ and represents the first example of a hexagonal structure among $5d$ oxides adopting a twin option. The sample shows spontaneous ferromagnetic magnetization below 430 K with a small saturation moment of 0.11 $\mu_B$/Os ion. This is the highest Curie temperature ($T_C$) reported for any bulk perovskite containing only $5d$ ions at the B site.

INTRODUCTION

Perovskites represent one of most robust and perhaps the most structurally variable classes of oxides.[1,2] The basic structure type $ABO_3$ allows for replacements at A and B sites in different ratios and for a wide variety of stacking sequences of the $AO_3$ dense packed layers, which results in a vast combinatorial diversity of individual configurations. As one of the consequences, this family of structures offers extensive opportunities for tuning the physical properties. Among the factors of influence, the Goldschmidt tolerance factor may serve as a quite reliable guide in anticipating distortions and stacking variants of any targeted perovskite structure.[3] The two ideal types of perovskites feature a cubic (*c*) or hexagonal (*h*) stacking sequence of the $AO_3$ layers, and virtually any combination of *c* and *h* sequences is possible.[2] In particular among the hexagonal perovskites, different combinatorial stacking of *c* and *h* layers leads to two different structure options: *shift* and *twin*, where the shift structures have a sequence of ($c_n hh$; n = number of layers) layers and twin structures are built up of ($c_n h$) layers.[2] The degrees of freedom mentioned give rise to interesting structure related properties such as microwave dielectrics (some recent examples are $La_5M_xTi_{4-x}O_{15}$, $Ba_8CoNb_6O_{24}$, $Ba_{10}Mg_{0.25}Ta_{7.9}O_{30}$ and $Ba_{10}Co_{0.25}Ta_{7.9}O_{30}$) and magnetic frustration (e.g. $Sr_2MIrO_6$; M = Ni and Zn, and $Sr_2MOsO_6$; M = Fe and Co).[4-9] Semiconducting or insulating magnetic perovskites (ferri- or ferromagnets) with high ordering temperatures are highly desirable for the use as spintronic devices. In this context, high ordering temperatures have been achieved recently in the double perovskites containing mixed 3*d*-4*d* or 5*d* metal ions at the B sites, e.g. $Sr_2FeMoO_6$ ($T_C$ =415 K), $Sr_2CrReO_6$ ($T_C$ =634 K), $Sr_2CrOsO_6$ ($T_C$ =725 K), and a quadruple perovskite $CaCu_3Fe_2Re_2O_{12}$ ($T_C$ = 560 K).[10-13] This is confirming that such materials offer a fertile ground to explore new perovskite-based materials with high transition temperatures. In our quest to explore materials with possible high magnetic transition temperatures, we have discovered an interesting 8-layered twinned hexagonal perovskite that contains only Os at the B-site and is exhibiting above room temperature ferromagnetic response ($T_C \sim 430$ K). So far, a twin option has not been observed in any $5d$ containing hexagonal perovskite.

EXPERIMENTAL SECTION

**Materials and synthesis.** Coarse crystalline samples of $Sr_8Os_{6.3}O_{24}$ were synthesized along a solid-state route. SrO (Aldrich, 99.9 %), $SrO_2$ (Aldrich, 99.9 %) and $OsO_2$ (Alfa Aesar, Os min 83 %) in the nominal ratio of 3:2:3 were heated at 800-1000º C for 48 hours in evacuated silica tubes. SrO and $SrO_2$ were mixed together and kept in one crucible whereas $OsO_2$ was put into a separate crucible positioned underneath the $SrO_x$ crucible. Due to the decomposition of $SrO_2$ above 650 K, $OsO_2$ converts to gaseous $OsO_4$ enabling reaction with the strontium oxides in the separate crucible. Prior to use, $OsO_2$ was fully oxidized by heating it in presence of $PbO_2$ (kept in separate crucibles) at 758 K for 48 h in an evacuated quartz tube. All the chemical manipulations were performed inside an Ar- filled glove box ($O_2$ and $H_2O$ < 1 ppm), except for quartz tube sealing. While heating, the tubes were always housed in a specially designed alumina casing which can be shut under static vacuum. This is to avoid any exposure of $OsO_4$ to the environment in case of silica tube rupture. Thick hexagonal crystals with tapered edges of length up to ~100 μm were obtained as major product when reaction was carried at 1273 K. The crystals were sonicated in ethanol for separation from the rest of the mass and air dried. The product appears to be stable in air for weeks.

*Caution: Tubes of sufficient thickness and length must be chosen in order to withstand the vapor pressure of $OsO_4$ during high temperature reaction. Since $OsO_4$, is highly toxic, silica tubes must be opened in a well-ventilated fume hood to avoid any potential contact with volatile $OsO_4$. Proper personal protective equipment must be worn while working with osmium outside the glove box.*

**Single crystal structure determination.** Single crystal X-ray diffraction (SCXRD) data sets were collected on an Apex-II CCD diffractometer (Bruker-AXS, Mo-Kα radiation, λ = 0.71073 Å, graphite monochromator) at ambient temperature using the APEX2 software suite.[14] A Lorentz and polarization factor and a numerical absorption corrections were applied. The structure was solved by charge flipping and the structure model was refined with SHELXL.[15,16] Crystallographic data, details of data collection and structure refinement, are given in Table 1 and atomic coordinates in Table 2. Selected bond lengths are given in Table 3. Crystallographic data have been deposited with Fachinformationszentrum Karlsruhe, D-76344 Eggenstein-Leopoldshafen (Germany) and can be obtained on quoting the depository number CSD-2059167.

**Powder X-ray diffraction.** Room temperature powder X-ray diffraction (PXRD) pattern in the 2θ range of 5-80º were recorded on a HUBER G670 imaging plate Guinier camera with Cu-Kα$_1$ radiation ($\lambda$ = 1.5406 Å). Unit cell constants and phase quantity were determined using Lebail fit and structure refinement was carried by Rietveld profile fit using the TOPAS-4.2.0.2 (Bruker-AXS) program.[17] The atomic positions obtained from the single crystal structure analysis were used as starting model. Refined parameter were cell constants, Wyckoff positions, occupation factor, scale factor, zero point of θ, sample displacement (mm), background as a Chebyshev polynomial of 20th degree, 1/x function, crystallite size and micro-strain. The structure was satisfactorily refined in the hexagonal *P*6$_3$*cm* space group with $R_{wp}$ = 4.2 % and GOF = 1.2, see Figure 1. Atomic positions from powder refinement are presented in Table S1 in SI.

**Magnetic property measurement.** The magnetization of two independent samples of carefully selected loose crystals held in pre-calibrated quartz tube holders was measured on a SQUID magnetometer (MPMS-XL7, Quantum Design). The temperature dependence of magnetization, *m*(*T*) was recorded during warming cycle after zero-field cooling (zfc) and during field-cooling (fc) under the applied magnetic fields $\mu_0H$ of 0.1, 3.5 and 7.0 T. The temperature range covered was 1.8-350 K and 300-750 K, without and with the instrument's oven insert, respectively. An isothermal magnetization curve was taken at $T = 1.8$ K. The susceptibility at infinite field was obtained using the simple Honda-Owen method. Before analyzing the data, the calculated sum of the diamagnetic increments for the compound (-512 × $10^{-6}$ emu mol$^{-1}$, using the average of the values for Os$^{4+}$ and Os$^{6+}$) was subtracted.[18]

RESULTS AND DISCUSSION

**Phase formation.** Black single crystals (Figure 1) with hexagonal habit were obtained in the course of solid-state reaction of SrO, SrO$_2$ and OsO$_2$ in vacuum at 1073-1273 K. The choice of the starting materials, their ratio, reaction temperature, purity of OsO$_2$, as well as the mode of reaction (direct reaction of SrO$_x$ and OsO$_2$ or in separate crucibles) played an important role in the phase formation. A ratio 3:2:3 of SrO, SrO$_2$ and OsO$_2$ at 1273 K resulted in the highest crystalline yield of the target phase whereas any deviation from this stoichiometry lead to higher amounts of side phases like Sr$_{1-x}$OsO$_3$, Sr$_2$OsO$_5$ and an unknown phase. At lower temperatures, smaller crystals were obtained making it difficult to separate manually. Use of as purchased OsO$_2$ and SrO$_2$ (in ratio 1:1 to 3:2) in separate crucibles lead to an almost equal mixture of crystals of cubic Sr$_x$OsO$_3$ and the title phase, hexagonal Sr$_8$Os$_{6.3}$O$_{24}$. At 1073 K only Sr$_x$OsO$_3$ had formed as a secondary phase along with the target phase, however, the size of the crystals of the latter was quite small. At 1273 K the crystals of the desired phase were sufficiently large and clean to be manually picked, but a minor amount of impurity Sr$_2$OsO$_5$ and a new unknown phase always kept sticking. A phase pure sample was not obtained, even after several trials with ratios of starting materials and at different temperatures. Best sample containing the highest amount of single crystals of appropriate size (~100 × 100 µm$^2$) and quantity sufficient for SCXRD, PXRD and magnetic measurements were obtained along the prescription as given in the experimental section.

The average composition of single crystals was determined using Scanning electron microscopy-energy dispersive X-ray analysis (SEM-EDX). An approximate Sr/Os ratio of ~ 1.27 was found for all the crystals (Figure S1 in Supplementary information), which resulted in the average composition of $Sr_8Os_{6.28}O_{24}$. The PXRD pattern of the powdered crystals was indexed and refined (LeBail method) in the primitive hexagonal cell with lattice constants, $a$ ~ 9.7 Å and $c$ ~ 18.2 Å, which was further corroborated by Laue diffraction of individual single crystals (Figure S2 in Supplementary information). The $c$–parameter ~ 18.2 Å pointed towards an 8–layered perovskite variant. Phase analysis was carried out by Rietveld refinement in the space group $P6_3cm$ using the CIF obtained from SCXRD. Apart from the title phase, a minor amount of impurity phases ($Sr_2OsO_5$ and an unknown phase) were also observed in the powder pattern. Since a quantitative refinement for a phase of unknown structure is not possible, a three phase LeBail fit on PXRD was performed to quantify the phases present, which resulted in the approximate phase percentages of ~2 % of $Sr_2OsO_5$ and ~8 % of an unknown trigonal phase ($a$ = 5.66 Å $c$ = 27.76 Å, possibly $R\bar{3}m$). The final fit shown in Figure 1, included in it is the LeBail fit for the unknown phase along with the Rietveld fit of the two identified phases. The compound starts decomposing above 1000 K and is only partially decomposed up to 1273 K with a weight loss of 5.7 % (Figure S3 in Supplementary information).

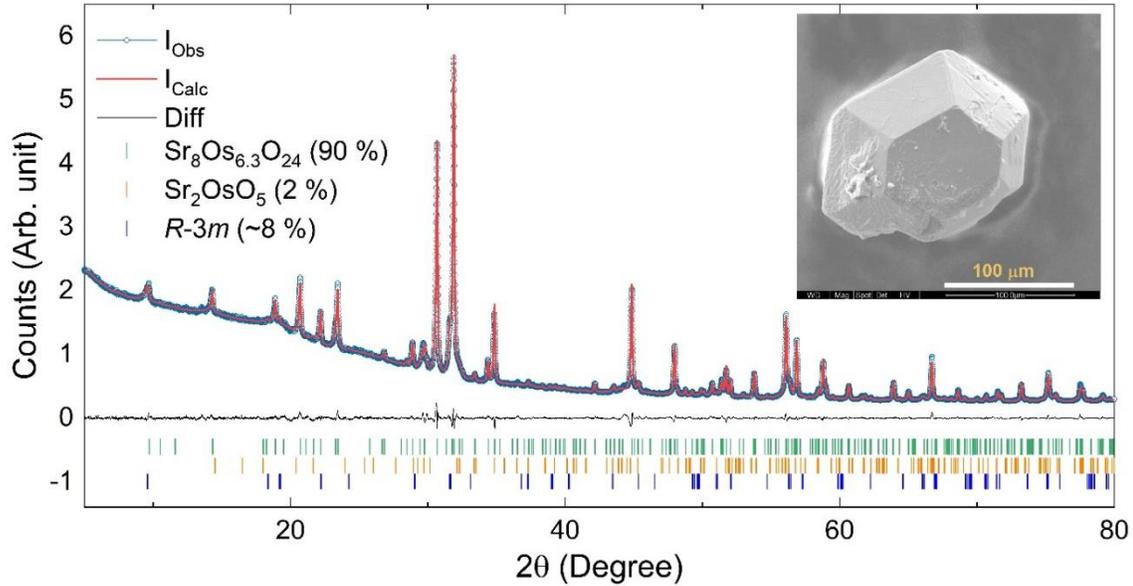

**Figure 1**. Room temperature PXRD data (blue circles) of $Sr_8Os_{6.3}O_{24}$ and its Rietveld fit (Red line). Black line is the difference curve and vertical bars are the allowed Bragg reflections for different phases. Inset show the SEM image of a typical single crystal on a 100-micron scale.

**Structure Analysis.** The crystal structure of $Sr_8Os_{6.3}O_{24}$ is shown in Figure 2. The compound crystallizes in an 8 layered hexagonal B-site deficient perovskite. The structure is isotypic to $Ba_8CuNb_6O_{24}$, $Ba_8Ga_{0.8}Ta_{5.92}O_{24}$ and $Ba_8Ti_3Nb_4O_{24}$ and is a twinned variant of a regular hexagonal perovskite with a tripled unit cell.[19-21] The 8-layers structure is built up from a particular stacking of cubic and hexagonal $SrO_3$ layers in the sequence of $(ccch)_2$, (*c* and *h* refer to cubic and hexagonal layers respectively). Such layered *twinned* structure requires only partial cation and vacancy ordering as opposed to the complete ordering in the *shift* types which have a stacking sequence of $(c_nhh)$. In the twinned structure two adjacent blocks of corner-shared octahedra (or $c_n$ layers) are in a twin relation to each other, when moving along the *c*-direction; here the *h* layer serves as the twin plane.[19, 22-23] The twinning results in expansion of the unit cell by a factor of √3 in both, *a* and *b* directions, compared to the 2H perovskite $BaNiO_3$, e.g. leading to tripling of the unit cell which can clearly be seen in the diffraction pattern. In the present 8H perovskite the cell enlargement is driven by a partial ordering of the $B$ sites in the layers: The Wyckoff site $2b$ in the layers at $z$ = 0.32 (and 0.82) remains unoccupied, whereas the $2a$ site at $z$ = 0.19 (and 0.69) is fully occupied (Os1) and the 4b sites (Os4 and Os5) are partially occupied. The $B$ sites in the $c$ layers (Os2, Os3, both on $6c$) are not affected by the occupational modulation. The five different B-sites for Os ions in this structure can, thus, be grouped into three corner-sharing octahedra (CSO; Os1, Os2 and Os3) and two face-sharing octahedra (FSO; Os4 and Os5). The presence of twinning is also implied by under occupation of FSO positions in the dimers and decreased periodicity. The metal-metal distance in the octahedral dimers ~ 2.432 Å supports this argument. These Os–Os distances are markedly shorter than those found in other hexagonal perovskites of the type $Ba_3MOs_2O_9$ (M = Li, Na, Cu and Zn) with Os–Os = 2.55–2.7 Å which feature fully occupied FSO sites.[24-25] From our refinements (single crystal and powder) we learnt that one Os site (Os4) is nearly fully occupied (93 %) while the other (Os5) is significantly under occupied (28%). From a crystal-chemical point of view, one would not expect a significant preference for a higher or lower degree of occupation for these two ((virtually) equivalent) sites on a first glance. The average Os5–O distances of 2.11 Å are larger than the average Os4–O distances (1.98 Å), but the closest Sr contacts are considerably reduced: Os5⋯Sr distances of 3.03 Å and 3.23 Å vs. Os4⋯Sr distances of > 3.4 Å. Considering, that the refined structure model in $P6_3cm$ is not fully satisfying as far as R-values and residual electron density, but also atomic displacement parameters are concerned, a structure model with reduced symmetry might point to a somewhat more sophisticated occupation pattern. Several models, considering space groups $P6_3$, $P\bar{3}c1$ or $P3c1$, e.g., have been tested, but a stable model resolving these issues could not be obtained; all these refinements suffer from large correlations. Similar problems were previously encountered during the superstructure refinement of a similar compound $Sr_4Ru_{3.05}O_{12}$ which were then partially resolved by electron diffraction studies where spots corresponding to superstructure were clearly seen.[26] A cation disorder in the occupancy of FSO was also implied by electron microscopy, which prevented an absolutely accurate structure solution from X-ray diffraction, a situation also suspected to have occurred in our case. Thus, we consider our structure model presented here as an averaged one, but nevertheless quite approximate to the real structure, as few preceding reports of compounds with very similar structural features lend strong support to the validity of our structure model.[19-21, 26] The resulting average composition $Sr_8Os_{6.3}O_{24}$, is also compliant with the other reported B-site deficient hexagonal perovskites of similar structures ($Ba_8Ga_{2.4}Ta_{4.96}O_{24}$ and $Sr_4Ru_{3.05}O_{12}$ = $Sr_8Ru_{6.1}O_{24}$) and is in excellent agreement with the EDX results.[20,26] These vacancies at the FSO sites helps to reduce the electrostatic repulsion between the short Os–Os contacts thus stabilising the 8-layer twinned structure. The CSO sites, on the other hand, are fully occupied and serve to link the face shared Os octahedral dimers to form a three−dimensional network. The Os–O bond distances lie in the range of 1.8–2.1 Å, which is commonly observed for Os(V) oxides.[27-30] The crystal data, final refined parameters obtained from single crystal data and important bond lengths are listed in table 1, 2 and 3 respectively. The anisotropic displacement parameters are listed in supplementary information (Table S2). The bond valence sum of the cations was calculated using the values reported by Gagné and Hawthorne.[31] BVS for Sr1, Sr4 and Sr5 is calculated to be 2.04, 2.07 and 2.09 in agreement with the divalent state, whereas the BVS for Sr2 and Sr3 is 2.35 and 2.47, might suggest a possible over-bonding. For Os1, Os2, Os3, Os4 and 5.6, 5.6, 5.2 and 4.9 which are acceptable for $Os^{5+}$ ions while for Os5, BVS of 3.7 clearly indicates the presence of significant vacancies.

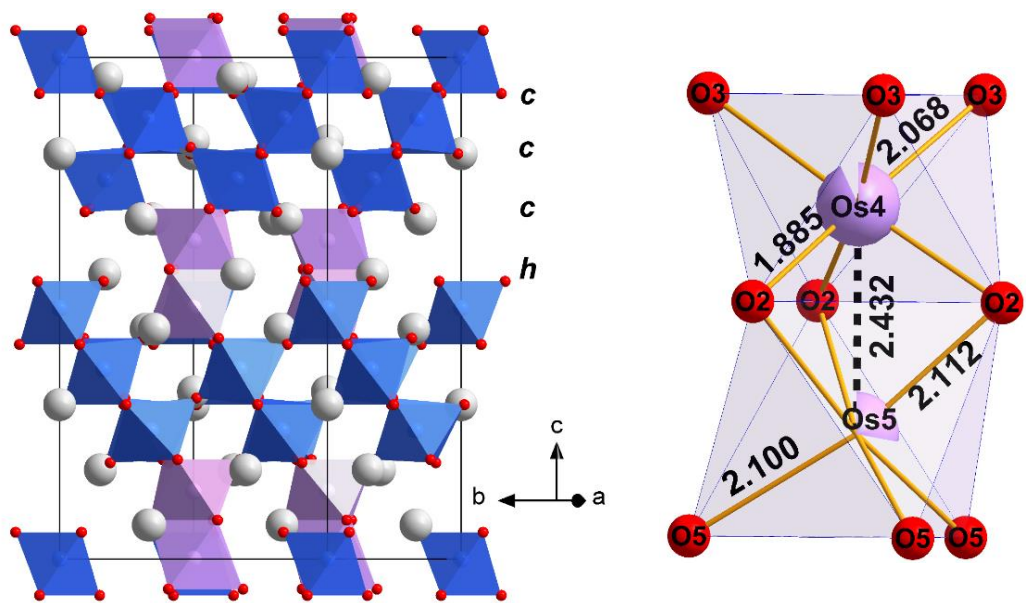

**Figure 2**: Crystal structure of 8-layered $Sr_8Os_{6.3}O_{24}$ (left) and coordination around Os atoms in the $Os_2O_9$ dimer (right). Os atoms with full and partial occupancies are represented as blue and purple spheres, respectively. Red and grey spheres are O and Sr atoms respectively. Bond distances are in Å.

**Table 1**. Crystal data and structure refinement for $Sr_8Os_{6.3}O_{24}$, lattice parameters in the parentheses were obtained from powder data.

| | |
|---|---|
| Empirical formula | $O_{24}$ $Os_{6.3}$ $Sr_8$ |
| Formula weight | 2283.39 |
| Temperature | 296(2) K |
| Wavelength | 0.71073 Å |
| Crystal system | Hexagonal |
| Space group | $P\,6_3\,c\,m$ |
| Unit cell dimensions | $a$ = 9.6988(3) Å (9.7065(3) Å) |
| | $c$ = 18.1657(5) Å (18.1852(3) Å) |
| Volume | 1479.85(10) Å$^3$ |
| Z | 3 |
| Density (calculated) | 7.673 g/cm$^3$ |
| Absorption coefficient | 61.793 mm$^{-1}$ |
| F(000) | 2920 |
| Crystal size | 0.044 x 0.031 x 0.029 mm$^3$ |

| | | |
|---|---|---|
| Theta range for data collection | 2.242 to 40.280°. | |
| Index ranges | -17<=h<=17, -17<=k<=17, -33<=l<=33 | |
| Reflections collected | 138418 | |
| Independent reflections | 3355 [R(int) = 0.0806] | |
| Completeness to theta = 25.242° | 99. 6 % | |
| Absorption correction | Numerical | |
| Max. and min. transmission | 0.3721 and 0.2313 | |
| Refinement method | Full-matrix least-squares on $F^2$ | |
| Data / restraints / parameters | 3355 / 1 / 74 | |
| Goodness-of-fit on $F^2$ | 1.067 | |
| Final R indices [I>2σ(I)] | R1 = 0.0333, wR2 = 0.0890 | |
| R indices (all data) | R1 = 0.0406, wR2 = 0.0919 | |
| Absolute structure parameter | 0.52(7) | |
| Extinction coefficient | 0.00045(5) | |
| Largest diff. peak and hole | 4.885 and -4.268 e.Å$^{-3}$ | |

**Table 2**. Atomic coordinates and equivalent isotropic displacement parameters (Å$^2$ × 10$^3$) for $Sr_8Os_{6.3}O_{24}$. U(eq) is defined as one third of the trace of the orthogonalized $U^{ij}$ tensor.

| Atom | Site | SOF | x | y | z | U(eq) |
|---|---|---|---|---|---|---|
| Os1 | 2a | 1 | 0 | 0 | 0.19212(5) | 2(1) |
| Os2 | 6c | 1 | 0 | -0.32625(6) | 0.07027(2) | 3(1) |
| Os3 | 6c | 1 | 0.34051(6) | 0.34051(6) | 0.44871(2) | 2(1) |
| Os4 | 4b | 0.931(5) | -0.3333 | 0.3333 | 0.32636 (4) | 5(1) |
| Os5 | 4b | 0.279(5) | -0.3333 | 0.3333 | 0.1925(1) | 2(1) |
| Sr1 | 4b | 1 | 0 | 0 | 0(3) | 13(1) |
| Sr2 | 6c | 1 | 0 | 0.3050(2) | 0.3702(2) | 11(1) |
| Sr3 | 6c | 1 | 0 | 0.3535(2) | 0.1506(2) | 26(1) |
| Sr4 | 6c | 1 | 0.3375(3) | 0.3375(3) | 0.2601(2) | 15(1) |
| Sr5 | 2a | 1 | -0.3333 | -0.6667 | 0.0255(2) | 10(1) |
| O1 | 6c | 1 | -0.164(2) | -0.164(2) | 0.246(1) | 31(1) |
| O2 | 12d | 1 | -0.177(2) | 0.321(2) | 0.2697(7) | 31(1) |
| O3 | 12d | 1 | 0.370(1) | 0.512(1) | 0.3904(6) | 31(1) |
| O4 | 6c | 1 | 0 | -0.150(2) | 0.1199(8) | 31(1) |
| O5 | 12d | 1 | 0.149(1) | -0.332(1) | 0.1324(8) | 31(1) |
| O6 | 6c | 1 | 0 | -0.488(2) | 0.0131(9) | 31(1) |
| O7 | 12d | 1 | 0.339(1) | 0.158(2) | 0.5000(7) | 31(1) |
| O8 | 6c | 1 | 0.189(2) | 0.189(2) | 0.384(1) | 31(1) |

**Table 3**. Important bond distances for $Sr_8Os_{6.3}O_{24}$.

| Bond | Distance (Å) | Bond | Distance (Å) |
|---|---|---|---|
| Os1-O1 (3×) | 1.86(2) | Os3-O6 (1×) | 2.029(17) |
| Os1-O4 (3×) | 1.956(17) | Os3-O8 (1×) | 1.879(17) |
| Os2-O4 (1×) | 1.936(17) | Os4-O2 (3×) | 1.885(14) |
| Os2-O5 (2×) | 1.856(14) | Os4-O3 (3×) | 2.067(11) |
| Os2-O6 (1×) | 1.885(17) | Os5-O2 (3×) | 2.112(14) |
| Os2-O7 (2×) | 1.946(14) | Os5-O5 (3×) | 2.100(15) |
| Os3-O3 (2×) | 1.867(11) | Os4-Os5 | 2.432(2) |
| Os3-O7 (2×) | 1.999(14) | | |

**Magnetization**. Temperature dependent magnetization data, $m(T)$, for one of the two samples are shown in Figure 3a for two applied fields. Both the curves rise sharply below 430 K (the temperature of a magnetic phase transition) and tend towards saturation below $\approx 200$ K. The temperature and field dependence below 430 K appears typical of a (weak) ferromagnetic material. The fc and zfc curve diverge only slightly at low temperature, indicating soft ferromagnetism and reasonably low magnetic disorder. The isothermal magnetization, $m(\mu_0H)$, at 1.8 K (Figure 3b) saturates at a low field of $\mu_0H \approx 0.02$ T and remains practically unchanged up to 7 T. The absence of a noticeable field hysteresis confirms the soft nature of the ferromagnetism. The saturated ferromagnetic moment per f. u. is small, only $m_{sat} \approx 0.8$ $\mu_B$, similarly low ferromagnetic saturation moments in the magnetically ordered state have been observed for $Ba_2NaOsO_6$ and $Ba_{11}Os_4O_{24}$.[32-34] The corrected inverse susceptibility $1/\chi(T)$ is shown in Figure 3c. Due to the low sample masses and the weak signal above 430 K an evaluation of the data is problematic. Fits of a Curie-Weiss law to data in temperature intervals of 500–750 K and alternatively in 550–750 K for the two samples result in effective magnetic moments ranging from 8 to 12 $\mu_B$ and Curie-Weiss parameters $\theta_{CW}$ from -1100 to -2200 K. In spite of the enormous inaccuracies of these parameters, this suggests an average magnetic moment of about 4 $\mu_B$ per Os species and the large negative $\theta_{CW}$ indicates (on average) strong antiferromagnetic interactions of the moments. The observed ferromagnetic signal then could be assigned to a weak ferromagnetic component of a basically canted (anti)ferromagnetic ordered structure or to a ferrimagnetic spin arrangement. Frustration of antiferromagnetic exchange interactions can be expected to play a role in the complex crystal structure of $Sr_8Os_{6.3}O_{24}$. The observed high ordering temperature is rare among semiconducting perovskites, with some double-perovskites as like $Sr_2FeMoO_6$ ($T_C$ = 415 K), $Sr_2CrReO_6$ ($T_C$ = 634 K), and $Sr_2CrOsO_6$ ($T_C$ = 725 K) and a quadruple perovskite $CaCu_3Fe_2Re_2O_{12}$ ($T_C$ = 560 K) serving as the only other examples.[10-13] However, all of them contain a combination of both 3$d$ and 4$d$ or 5$d$ elements at the B sites, thus a high transition temperature is naturally expected.

Surprisingly, very recently in thin films of a cubic double perovskite $Sr_3OsO_6$ which contains only Os ion at the B-sites, extremely high $T_C$ of ~1064 K was reported which is the highest among any known semiconducting oxide.[35] However, the bulk sample lacks any magnetic order even down to 2 K.[36] Apparently, $Sr_8Os_{6.3}O_{24}$ has the highest ordering temperature (430 K) among bulk perovskite compounds containing only 5$d$ metal at B sites. Moreover it is the first example of an 8-layered hexagonal structure in 5$d$ oxides exhibiting a twin option.

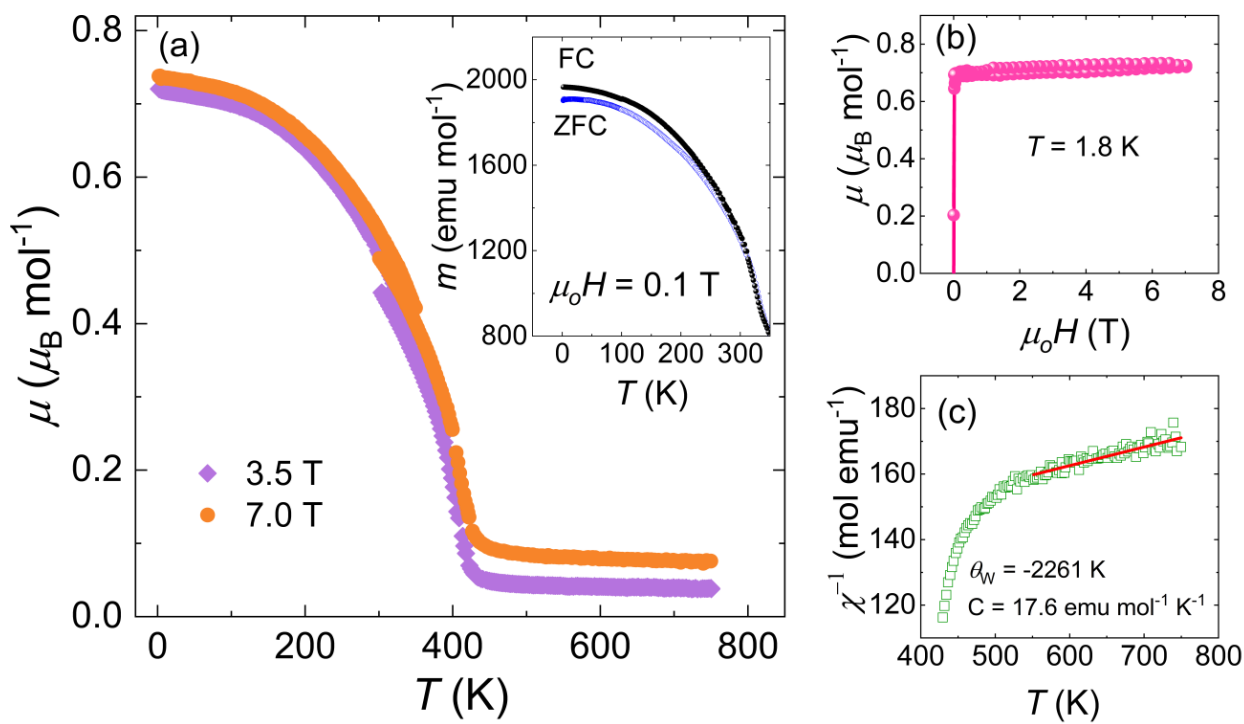

**Figure 3**: (a) Temperature dependence of magnetization for two applied fields, (b) isothermal magnetization at 1.8 K and (c) inverse corrected susceptibility along with a fit to the Curie-Weiss equation for $Sr_8Os_{6.3}O_{24}$. The inset of (a) shows the zfc and fc protocol magnetization data.

CONCLUSION

Single crystals synthesis of a new 8-layered twinned hexagonal perovskite $Sr_8Os_{6.3}O_{24}$, with significant B-site deficiency is reported. It is the first $5d$ perovskite to crystallize in 8-layer twinned hexagonal structure. It possesses a high ferromagnetic ordering temperature of 430 K. This is apparently the highest ordering temperature featuring a (weak) ferromagnetic component for a bulk perovskite containing only $5d$ metal at B-site.

ASSOCIATED CONTENT

***Supporting Information**

SEM-EDX analysis, Laue diffraction, thermal decomposition profile and tables for crystallographic parameters obtained from powder refinement, and anisotropic displacement parameters from single crystal. The following files are available free of charge.

AUTHOR INFORMATION

**Corresponding Author**

Martin Jansen- *Max Planck Institute for Solids State Research, Heisenberg Str. 1, 70569 Stuttgart, Germany;* Email: m.jansen@fkf.mpg.de

**Author Contributions**

All authors have given approval to the final version of the manuscript

ACKNOWLEDGEMENT

The authors thank Maximilian Knies for single crystal data collection. GST thank the Cluster of Excellence *ct.qmat* (EXC 2147) funded by the DFG for partial support. The work at TUD was financially supported by the Deutsche Forschungsgemeinschaft (DFG) within the SFB 1143 "Correlated Magnetism–From Frustration to Topology" (project-id 247310070).

REFERENCES

1. Vasala, S.; Karppinen, M. A$_2$B′B″O$_6$ perovskites: a review. *Prog. Solid State Chem*. 2015, *43*, 1-36.

2. Tilley, R. J. D. *Pervoskites: Structure–Property Relationships;* John Wiley & Sons, Ltd: United Kingdom, **2016**.

3. Goldschmidt, V. M. Die Gesetze der Krystallochemie. *Sci. Nat.* **1926**, *14*, 477-485.

4. Kuang, X.; Allix, M. M. B.; Claridge, J. B.; Niu, H. J.; Rosseinsky, M. J.; Ibberson, R. M.; Iddles, D. M. Crystal structure, microwave dielectric properties and AC conductivity of B-cation deficient hexagonal perovskites La$_5$M$_x$Ti$_{4-x}$O$_{15}$ ($x$ = 0.5, 1; M = Zn, Mg, Ga, Al). *J. Mater. Chem*. **2006**, *16*, 1038−1045.

5. Mallinson, P. M.; Allix, M. M. B.; Claridge, J. B.; Ibberson, R. M.; Iddles, D. M.; Price, T.; Rosseinsky, M. J. Ba$_8$CoNb$_6$O$_{24}$: A $d^0$ dielectric oxide host containing ordered d$^7$ cation layers 1.88 nm apart. *Angew. Chem., Int. Ed*. **2005**, *117*, 7911−7914.

6. Mallinson, P.; Claridge, J. B.; Iddles, D.; Price, T.; Ibberson, R. M.; Allix, M.; Rosseinsky, M. New 10-layer hexagonal perovskites: Relationship between cation and vacancy ordering and microwave dielectric loss. J. *Chem. Mater*. **2006**, *18*, 6227–6238.

7. Ou, X.; Li, Z.; Fan, F.; Wang, H.; Wu, H. Long-range magnetic interaction and frustration in double perovskites Sr$_2$NiIrO$_6$ and Sr$_2$ZnIrO$_6$. *Scientific Reports* **2014**, *4*, 7542.

8. Paul, A. K.; Reehuis, M.; Ksenofontov, V.; Yan, B.; Hoser A.; Többens, D. M.; Abdala, P. M.; Adler, P.; Jansen, M.; Felser, C. Lattice instability and competing spin structures in the double perovskite insulator Sr$_2$FeOsO$_6$. *Phys. Rev. Lett*. **2013**, *111*, 167205.

9. Yan, B.; Paul, A. K.; Kanungo, S.; Reehuis, M.; Hoser, A.; Többens, D. M.; Schnelle, W.; Williams, R. C.; Lancaster, T.; Xiao, F.; Möller, J. S.; Blundell, S. J.; Hayes, W.; Felser, C.; Jansen, M. Lattice-site-specific spin dynamics in double perovskite Sr$_2$CoOsO$_6$. *Phys. Rev. Lett*. **2014**, *112*, 147202.

10. Kobayashi, K.-I.; Kimura, T.; Sawada, H.; Terakura, K.; Tokura, Y. Room-temperature magnetoresistance in an oxide material with an ordered double-perovskite structure. *Nature* **1998**, *395*, 677-680.

11. Moussa, S. M.; Claridge, J. B.; Rosseinsky, M. J.; Clarke, S.; Ibberson R. M.; Price T.; Iddles, D. M.; Sinclair, D. C. Metallic ordered double-perovskite Sr$_2$CrReO$_6$ with maximal Curie temperature of 635 K. *Appl. Phys. Lett*. **2002**, *81*, 328-330.

12. Krockenberger, Y.; Mogare, K.; Reehuis, M.; Tovar, M.; Jansen, M.; Vaitheeswaran, G.; Kanchana, V.; Bultmark, F.; Delin, A.; Wilhelm, F.; Rogalev, A.; Winkler, A; Alff, L. $Sr_2CrOsO_6$: End point of a spin-polarized metal-insulator transition by $5d$ band filling. *Phys. Rev. B* **2007**, *75*, 020404(R).

13. Chen, W.; Mizumaki, M.; Seki, H.; Senn, M. S.; Saito, T.; Kan, D.; Attfield J. P.; Shimakawa Y. A half-metallic A- and B-site-ordered quadruple perovskite oxide $CaCu_3Fe_2Re_2O_{12}$ with large magnetization and a high transition temperature. *Nat. Commun*. **2014**, *5*, 3909.

14. APEX2, Version 2014/9, Bruker AXS Inc., Madison, Wisconsin, USA, **2014**.

15. Oszlányi, G.; Sütő, A. Ab initio structure solution by charge flipping. *Acta Cryst.* **2004**, *A60*, 134-141 *ibid*. The charge flipping algorithm. **2008**, *A64*, 123-134.

16. Sheldrick, G. M. SHELXL- Integrated space-group and crystal-structure determination. *Acta Cryst*. C, **2015**, *71*, 3-8.

17. TOPAS-V4.2.0.2: General Profile and Structure Analysis Software for Powder Diffraction Data; Bruker AXS GmbH: Karlsruhe, Germany, **2018**.

18. Bain, G. A.; Berry, J. F. Diamagnetic Corrections and Pascal's Constants. *J. Chem. Education,* **2008**, *85*, 532-536.

19. Yin, C.; Tao, F.; Zhao, L.; Wang, X.; Ming, X.; Liang, C.; Kuang, X. Shift–Twin option in eight-layer hexagonal perovskite niobates $Ba_8MNb_6O_{24}$. *Inorg. Chem*. **2020**, *59*, 16375−16384.

20. Cao, J.; Kuang, X.; Allix, M.; Dickinson, C.; Claridge, J. B.; Rosseinsky, M. J.; Iddles, D. M.; Su, Q. New 8-layer twinned hexagonal perovskite microwave dielectric ceramics $Ba_8Ga_{4-x}Ta_{4+0.6x}O_{24}$. *Chem. Mater*. **2011**, *23*, 5058−5067.

21. Teneze, N.; Boullay, P.; Petricek, V.; Trolliard, G.; Mercurio, D. Structural study of the cation ordering in the ternary oxide $Ba_8Ti_3Nb_4O_{24}$, *Solid State Sci.* **2002**, *4*, 1129–1136.

22. Moussa, S. M.; Claridge, J. B.; Rosseinsky, M. J.; Clarke, S.; Ibberson, R. M.; Price, T.; Iddles, D. M.; Sinclair, D. C. $Ba_8ZnTa_6O_{24}$: a high-Q microwave dielectric from a potentially diverse homologous series. *Appl. Phys. Lett.* **2003**, *82*, 4537−4539.

23. Abakumov, A. M.; Tendeloo, G. V.; Scheglov, A. A.; Shpanchenko, R. V.; Antipov, E. V. The crystal structure of $Ba_8Ta_6NiO_{24}$: Cation Ordering in Hexagonal Perovskites. *J. Solid State Chem.* **1996**, *125*, 102−107.

24. Feng, H. L.; and Jansen, M. $Ba_3CuOs_2O_9$ and $Ba_3ZnOs_2O_9$, a comparative study. *J. Solid State Chem.* **2018**, *258*, 776–780.

25. Stitzer, K. E.; El Abed. A.; Smith, M. D.; Davis, M. J.; Kim, S.-J.; Darriet, J.; zur Loye, H.-C. Crystal growth of novel osmium-containing triple perovskites. *Inorg. Chem.*, **2003**, *42*, 947-949.

26. Renard, C.; Daviero-Minaud, S.; Huve, M.; Abraham, F. $Sr_4Ru_{3.05}O_{12}$: A new member of the hexagonal perovskite family. *J. Solid State Chem.* **1999**, *144*, 125−135.

27. Thakur, G. S.; Feng, H. L.; Schnelle, W.; Felser, C.; Jansen, M. Structure and magnetism of new A-and B-site ordered double perovskites $ALaCuOsO_6$ (A= Ba and Sr). *J. Solid State Chem.* **2021**, *293*, 121784.

28. Paul, A. K.; Jansen, M.; Yan, B.; Felser, C.; Reehuis, M.; Abdala, P. M. Synthesis, Crystal Structure, and physical properties of $Sr_2FeOsO_6$. *Inorg. Chem.* **2013**, *52*, 6713–6719.

29. Feng, H. L.; Schnelle, W.; Tjeng, L. H.; Jansen, M. Synthesis, crystal structures, and magnetic properties of double perovskites $SrLaNiOsO_6$ and $BaLaNiOsO_6$. *Solid State Commun.* **2016**, *243*, 49–54.

30. Kermarrec, E.; Marjerrison, C. A.; Thompson, C. M.; Maharaj, D. D.; Levin, K.; Kroeker, S.; Granroth, G. E.; Flacau, R.; Yamani, Z.; Greedan, J. E.; Gaulin, B. D. Frustrated fcc antiferromagnet $Ba_2YOsO_6$: Structural characterization, magnetic properties, and neutron scattering studies *Phys. Rev. B,* **2015**, *91*, 075133.

31. Gagné, O. C.; and Hawthorne, F. C. Comprehensive derivation of bond-valence parameters for ion pairs involving oxygen. *Acta Cryst*. **2015**. *B71*, 562–578.

32. Stitzer, K. E.; Smith, M. D. & zur Loye, H. C. Crystal growth of $Ba_2MOsO_6$ (M= Li, Na) from reactive hydroxide fluxes. *Solid State Sci*. **2002**, *4*, 311-316.

33. Erickson, A. S.; Misra, S.; Miller, G. J.; Gupta, R. R; Schlesinger, Z.; Harrison, W. A.; Kim, J. M.; Fisher, I. R. Ferromagnetism in the Mott Insulator $Ba_2NaOsO_6$. *Phys. Rev. Lett.* **2007**, *99*, 016404.

34. Wakeshima, M.; Hinatsu, Y. Crystal structure and magnetic ordering of novel perovskite-related barium osmate $Ba_{11}Os_4O_{24}$. *Solid State Commun*. **2005**, *136*, 499–503.

35. Wakabayashi, Y. K.; Krockenberger, Y.; Tsujimoto, N.; Boykin, T.; Tsuneyuki, S.; Taniyasu, Y.; Yamamoto, H. Ferromagnetism above 1000 K in a highly cation-ordered double-perovskite insulator $Sr_3OsO_6$. *Nat Commun*. **2019**, *10*, 535.

36. Chen, J.; Feng, H. L.; Matsushita, Y.; Belik, A. A.; Tsujimoto, Y.; Tanaka, M.; Chung, D. Y.; Yamaura, K. Study of polycrystalline bulk $Sr_3OsO_6$ double-perovskite insulator: comparison with 1000 K ferromagnetic epitaxial films. *Inorg. Chem*. **2020**, *59*, 4049–4057.